\documentclass[twocolumn,letterpaper,aps,prc,longbibliography,superscriptaddress,nofootinbib,floatfix]{revtex4-1}

\usepackage{multirow}
\usepackage{amsmath}
\usepackage{graphicx}	
\usepackage{xspace}	
\usepackage{longtable}
\newcommand{\pt}{\mbox{$p_T$}\xspace}
\newcommand{\piz}{\mbox{$\pi^0$}\xspace}
\newcommand{\gevc}{\mbox{GeV/$c$}\xspace}
\newcommand{\raa}{\mbox{$R_{AA}$}\xspace}

\newcommand{\Npart}{\mbox{$N_{\rm part}$}\xspace}
\newcommand{\Ncoll}{\mbox{$N_{\rm coll}$}\xspace}

\newcommand{\sqsn}{\mbox{$\sqrt{s_{_{NN}}}$}\xspace}

\newcommand{\pp}{\mbox{$p$$+$$p$}\xspace}

\newcommand{\auau}{\mbox{Au$+$Au}\xspace}

\newcommand{\uu}{\mbox{U$+$U}\xspace}
\newcommand{\aaa}{\mbox{$A$$+$$A$}\xspace}

\begin{document}

\title{Production of $\pi^0$ and $\eta$ mesons in U$+$U 
collisions at $\sqrt{s_{_{NN}}}=192$ GeV}

\newcommand{\abilene}{Abilene Christian University, Abilene, Texas 79699, USA}
\newcommand{\augie}{Department of Physics, Augustana University, Sioux Falls, South Dakota 57197, USA}
\newcommand{\banaras}{Department of Physics, Banaras Hindu University, Varanasi 221005, India}
\newcommand{\barc}{Bhabha Atomic Research Centre, Bombay 400 085, India}
\newcommand{\baruch}{Baruch College, City University of New York, New York, New York, 10010 USA}
\newcommand{\bnlcoll}{Collider-Accelerator Department, Brookhaven National Laboratory, Upton, New York 11973-5000, USA}
\newcommand{\bnlphys}{Physics Department, Brookhaven National Laboratory, Upton, New York 11973-5000, USA}
\newcommand{\caucr}{University of California-Riverside, Riverside, California 92521, USA}
\newcommand{\charlesczech}{Charles University, Ovocn\'{y} trh 5, Praha 1, 116 36, Prague, Czech Republic}
\newcommand{\ciae}{Science and Technology on Nuclear Data Laboratory, China Institute of Atomic Energy, Beijing 102413, People's Republic of China}
\newcommand{\cns}{Center for Nuclear Study, Graduate School of Science, University of Tokyo, 7-3-1 Hongo, Bunkyo, Tokyo 113-0033, Japan}
\newcommand{\colorado}{University of Colorado, Boulder, Colorado 80309, USA}
\newcommand{\columbia}{Columbia University, New York, New York 10027 and Nevis Laboratories, Irvington, New York 10533, USA}
\newcommand{\czechtech}{Czech Technical University, Zikova 4, 166 36 Prague 6, Czech Republic}
\newcommand{\debrecen}{Debrecen University, H-4010 Debrecen, Egyetem t{\'e}r 1, Hungary}
\newcommand{\elte}{ELTE, E{\"o}tv{\"o}s Lor{\'a}nd University, H-1117 Budapest, P{\'a}zm{\'a}ny P.~s.~1/A, Hungary}
\newcommand{\eszterhazy}{Eszterh\'azy K\'aroly University, K\'aroly R\'obert Campus, H-3200 Gy\"ongy\"os, M\'atrai \'ut 36, Hungary}
\newcommand{\ewha}{Ewha Womans University, Seoul 120-750, Korea}
\newcommand{\famu}{Florida A\&M University, Tallahassee, FL 32307, USA}
\newcommand{\fsu}{Florida State University, Tallahassee, Florida 32306, USA}
\newcommand{\gsu}{Georgia State University, Atlanta, Georgia 30303, USA}
\newcommand{\hanyang}{Hanyang University, Seoul 133-792, Korea}
\newcommand{\hiroshima}{Hiroshima University, Kagamiyama, Higashi-Hiroshima 739-8526, Japan}
\newcommand{\ihepprot}{IHEP Protvino, State Research Center of Russian Federation, Institute for High Energy Physics, Protvino, 142281, Russia}
\newcommand{\illuiuc}{University of Illinois at Urbana-Champaign, Urbana, Illinois 61801, USA}
\newcommand{\inrras}{Institute for Nuclear Research of the Russian Academy of Sciences, prospekt 60-letiya Oktyabrya 7a, Moscow 117312, Russia}
\newcommand{\instpasczech}{Institute of Physics, Academy of Sciences of the Czech Republic, Na Slovance 2, 182 21 Prague 8, Czech Republic}
\newcommand{\isu}{Iowa State University, Ames, Iowa 50011, USA}
\newcommand{\jaea}{Advanced Science Research Center, Japan Atomic Energy Agency, 2-4 Shirakata Shirane, Tokai-mura, Naka-gun, Ibaraki-ken 319-1195, Japan}
\newcommand{\jeonbuk}{Jeonbuk National University, Jeonju, 54896, Korea}
\newcommand{\jyvaskyla}{Helsinki Institute of Physics and University of Jyv{\"a}skyl{\"a}, P.O.Box 35, FI-40014 Jyv{\"a}skyl{\"a}, Finland}
\newcommand{\kek}{KEK, High Energy Accelerator Research Organization, Tsukuba, Ibaraki 305-0801, Japan}
\newcommand{\korea}{Korea University, Seoul, 02841}
\newcommand{\kurchatov}{National Research Center ``Kurchatov Institute", Moscow, 123098 Russia}
\newcommand{\kyoto}{Kyoto University, Kyoto 606-8502, Japan}
\newcommand{\labllr}{Laboratoire Leprince-Ringuet, Ecole Polytechnique, CNRS-IN2P3, Route de Saclay, F-91128, Palaiseau, France}
\newcommand{\lahorelums}{Physics Department, Lahore University of Management Sciences, Lahore 54792, Pakistan}
\newcommand{\lawllnl}{Lawrence Livermore National Laboratory, Livermore, California 94550, USA}
\newcommand{\losalamos}{Los Alamos National Laboratory, Los Alamos, New Mexico 87545, USA}
\newcommand{\lund}{Department of Physics, Lund University, Box 118, SE-221 00 Lund, Sweden}
\newcommand{\maryland}{University of Maryland, College Park, Maryland 20742, USA}
\newcommand{\mass}{Department of Physics, University of Massachusetts, Amherst, Massachusetts 01003-9337, USA}
\newcommand{\michigan}{Department of Physics, University of Michigan, Ann Arbor, Michigan 48109-1040, USA}
\newcommand{\muhlenberg}{Muhlenberg College, Allentown, Pennsylvania 18104-5586, USA}
\newcommand{\myongji}{Myongji University, Yongin, Kyonggido 449-728, Korea}
\newcommand{\nagasaki}{Nagasaki Institute of Applied Science, Nagasaki-shi, Nagasaki 851-0193, Japan}
\newcommand{\nara}{Nara Women's University, Kita-uoya Nishi-machi Nara 630-8506, Japan}
\newcommand{\natmephi}{National Research Nuclear University, MEPhI, Moscow Engineering Physics Institute, Moscow, 115409, Russia}
\newcommand{\newmex}{University of New Mexico, Albuquerque, New Mexico 87131, USA}
\newcommand{\nmsu}{New Mexico State University, Las Cruces, New Mexico 88003, USA}
\newcommand{\northcg}{Physics and Astronomy Department, University of North Carolina at Greensboro, Greensboro, North Carolina 27412, USA}
\newcommand{\ohio}{Department of Physics and Astronomy, Ohio University, Athens, Ohio 45701, USA}
\newcommand{\ornl}{Oak Ridge National Laboratory, Oak Ridge, Tennessee 37831, USA}
\newcommand{\orsay}{IPN-Orsay, Univ.~Paris-Sud, CNRS/IN2P3, Universit\'e Paris-Saclay, BP1, F-91406, Orsay, France}
\newcommand{\pnpi}{PNPI, Petersburg Nuclear Physics Institute, Gatchina, Leningrad region, 188300, Russia}
\newcommand{\pusan}{Pusan National University, Pusan 46241, Korea}
\newcommand{\riken}{RIKEN Nishina Center for Accelerator-Based Science, Wako, Saitama 351-0198, Japan}
\newcommand{\rikjrbrc}{RIKEN BNL Research Center, Brookhaven National Laboratory, Upton, New York 11973-5000, USA}
\newcommand{\rikkyo}{Physics Department, Rikkyo University, 3-34-1 Nishi-Ikebukuro, Toshima, Tokyo 171-8501, Japan}
\newcommand{\saispbstu}{Saint Petersburg State Polytechnic University, St.~Petersburg, 195251 Russia}
\newcommand{\seoulnat}{Department of Physics and Astronomy, Seoul National University, Seoul 151-742, Korea}
\newcommand{\stonybrkc}{Chemistry Department, Stony Brook University, SUNY, Stony Brook, New York 11794-3400, USA}
\newcommand{\stonycrkp}{Department of Physics and Astronomy, Stony Brook University, SUNY, Stony Brook, New York 11794-3800, USA}
\newcommand{\sungskku}{Sungkyunkwan University, Suwon, 440-746, Korea}
\newcommand{\tenn}{University of Tennessee, Knoxville, Tennessee 37996, USA}
\newcommand{\titech}{Department of Physics, Tokyo Institute of Technology, Oh-okayama, Meguro, Tokyo 152-8551, Japan}
\newcommand{\tsukuba}{Tomonaga Center for the History of the Universe, University of Tsukuba, Tsukuba, Ibaraki 305, Japan}
\newcommand{\vandy}{Vanderbilt University, Nashville, Tennessee 37235, USA}
\newcommand{\weizmann}{Weizmann Institute, Rehovot 76100, Israel}
\newcommand{\wigner}{Institute for Particle and Nuclear Physics, Wigner Research Centre for Physics, Hungarian Academy of Sciences (Wigner RCP, RMKI) H-1525 Budapest 114, POBox 49, Budapest, Hungary}
\newcommand{\yonsei}{Yonsei University, IPAP, Seoul 120-749, Korea}
\newcommand{\zagreb}{Department of Physics, Faculty of Science, University of Zagreb, Bijeni\v{c}ka c.~32 HR-10002 Zagreb, Croatia}
\affiliation{\abilene}
\affiliation{\augie}
\affiliation{\banaras}
\affiliation{\barc}
\affiliation{\baruch}
\affiliation{\bnlcoll}
\affiliation{\bnlphys}
\affiliation{\caucr}
\affiliation{\charlesczech}
\affiliation{\ciae}
\affiliation{\cns}
\affiliation{\colorado}
\affiliation{\columbia}
\affiliation{\czechtech}
\affiliation{\debrecen}
\affiliation{\elte}
\affiliation{\eszterhazy}
\affiliation{\ewha}
\affiliation{\famu}
\affiliation{\fsu}
\affiliation{\gsu}
\affiliation{\hanyang}
\affiliation{\hiroshima}
\affiliation{\ihepprot}
\affiliation{\illuiuc}
\affiliation{\inrras}
\affiliation{\instpasczech}
\affiliation{\isu}
\affiliation{\jaea}
\affiliation{\jeonbuk}
\affiliation{\jyvaskyla}
\affiliation{\kek}
\affiliation{\korea}
\affiliation{\kurchatov}
\affiliation{\kyoto}
\affiliation{\labllr}
\affiliation{\lahorelums}
\affiliation{\lawllnl}
\affiliation{\losalamos}
\affiliation{\lund}
\affiliation{\maryland}
\affiliation{\mass}
\affiliation{\michigan}
\affiliation{\muhlenberg}
\affiliation{\myongji}
\affiliation{\nagasaki}
\affiliation{\nara}
\affiliation{\natmephi}
\affiliation{\newmex}
\affiliation{\nmsu}
\affiliation{\northcg}
\affiliation{\ohio}
\affiliation{\ornl}
\affiliation{\orsay}
\affiliation{\pnpi}
\affiliation{\pusan}
\affiliation{\riken}
\affiliation{\rikjrbrc}
\affiliation{\rikkyo}
\affiliation{\saispbstu}
\affiliation{\seoulnat}
\affiliation{\stonybrkc}
\affiliation{\stonycrkp}
\affiliation{\sungskku}
\affiliation{\tenn}
\affiliation{\titech}
\affiliation{\tsukuba}
\affiliation{\vandy}
\affiliation{\weizmann}
\affiliation{\wigner}
\affiliation{\yonsei}
\affiliation{\zagreb}
\author{U.~Acharya} \affiliation{\gsu} 
\author{C.~Aidala} \affiliation{\losalamos} \affiliation{\michigan} 
\author{N.N.~Ajitanand} \altaffiliation{Deceased} \affiliation{\stonybrkc} 
\author{Y.~Akiba} \email[PHENIX Spokesperson: ]{akiba@rcf.rhic.bnl.gov} \affiliation{\riken} \affiliation{\rikjrbrc} 
\author{R.~Akimoto} \affiliation{\cns} 
\author{J.~Alexander} \affiliation{\stonybrkc} 
\author{K.~Aoki} \affiliation{\kek} \affiliation{\riken} 
\author{N.~Apadula} \affiliation{\isu} \affiliation{\stonycrkp} 
\author{H.~Asano} \affiliation{\kyoto} \affiliation{\riken} 
\author{E.T.~Atomssa} \affiliation{\stonycrkp} 
% ...
\collaboration{PHENIX Collaboration} \noaffiliation

\date{\today}


\begin{abstract}


The PHENIX experiment at the Relativistic Heavy Ion Collider measured 
$\pi^0$ and $\eta$ mesons at midrapidity in U$+$U collisions at 
$\sqrt{s_{_{NN}}}=192$ GeV in a wide transverse momentum range. 
Measurements were performed in the 
$\pi^0(\eta)\rightarrow\gamma\gamma$ decay modes.  A strong 
suppression of $\pi^0$ and $\eta$ meson production at high transverse 
momentum was observed in central U$+$U collisions relative to binary 
scaled $p$$+$$p$ results. Yields of $\pi^0$ and $\eta$ mesons measured 
in U$+$U collisions show similar suppression pattern to the ones 
measured in Au$+$Au collisions at $\sqrt{s_{_{NN}}}=200$ GeV for similar 
numbers of participant nucleons. The $\eta$/$\pi^0$ ratios do not show 
dependence on centrality or transverse momentum, and are consistent with 
previously measured values in hadron-hadron, hadron-nucleus, 
nucleus-nucleus, and $e^+e^-$ collisions.

\end{abstract}

\maketitle

\section{Introduction}

Extensive studies of heavy-ion collisions (A$+$A) at the Relativistic 
Heavy Ion Collider (RHIC) resulted in the discovery of the quark-gluon 
plasma (QGP)~\cite{Arsene:2004fa,Back:2004je,Adams:2005dq,Adcox:2004mh}. 
Subsequent measurements at the Large Hadron 
Collider~\cite{CMS:2012aa,Abelev:2012hxa,Aad:2012vca,Connors:2017ptx} 
confirmed the suppression of high \pt hadrons characteristic of the QGP 
and firmly established the existence of true jet quenching.  Since then, 
one of the main efforts of RHIC experiments was directed towards 
detailed studies of the properties of the new state of nuclear matter, 
in part by making more differential and more precise measurements, but 
also by varying the collision energy and system size.  The culmination 
of the latter was colliding \uu, the largest ever nucleus-nucleus 
collision system studied so far at RHIC or the Large Hadron Collider.

Creation of the QGP causes a variety of observable effects, including 
the so-called 
jet-quenching~\cite{Bjorken:1982tu,Baier:2000mf,Wang:1994fx}, which 
manifests itself by strongly suppressed production of high transverse 
momentum (\pt) hadrons in A$+$A, relative to the yields measured in 
proton-proton ($p$$+$$p$) collisions and scaled by the number of expected 
binary nucleon-nucleon collisions.  The suppression is related to the 
energy loss of hard-scattered partons in a quark-gluon medium via 
bremsstrahlung and elastic scatterings. Parton energy loss is 
characterized by the $\hat{q}$ transport parameter, which represents the 
squared four-momentum transfer between the parton and the medium per 
unit path length and carries information on the medium 
coupling~\cite{Baier:2000mf, Majumder:2007zh}. Values of the $\hat{q}$ 
parameter cannot yet be estimated from first principles. Instead, 
several phenomenological jet-quenching 
models~\cite{Qin:2007rn,Chen:2011vt,Young:2011ug,Majumder:2011uk,Xu:2014ica} 
exist, all based on experimental results.

Quantitatively, medium effects in A$+$A are usually characterized with 
the nuclear modification factor ($\raa$):
\begin{equation}
   R_{AA}^{\rm cent}(\pt) = \frac{1}{T_{AA}^{\rm cent}}\frac{dN_{AA}^{\rm cent}/d\pt}{d\sigma_{pp}/d\pt},
\end{equation}
where $dN_{AA}^{\rm cent}/d\pt$ is the particle yield measured in A+A 
collisions for a given centrality class (cent), $d\sigma_{pp}$/$p_T$ 
is the particle production cross section measured in $\pp$ collisions at 
the same collision energy while $T_{AA}^{\rm cent}$ is the nuclear 
thickness function for the event centrality class~\cite{Miller:2007ri}.

Measurements of the production of different types of mesons 
allow a systematic study of jet quenching with respect to 
the fragmentation function and quantum numbers (mass, flavor, spin, etc.) of the final state hadrons. For example, $\pi^0$ mesons contain only the first generation quarks ($u$, $d$) and thus are produced abundantly, while $\eta$ mesons have a hidden strangeness content and four times larger mass than $\pi^0$.
Measurement of the $\eta/\pi^0$ 
ratios in $\aaa$ gives an opportunity to better understand the possible changes of parton 
fragmentation mechanisms with respect to system size, collision energy 
and geometry. They are also an important input for the measurement of 
direct photons.

In this paper we present results on $\pi^0$ and $\eta$ meson $p_T$ 
invariant yields, \raa, and $\eta/\pi^0$ ratios in \uu collisions at 
$\sqrt{s_{_{NN}}}=192$ GeV. The $^{238}$U+$^{238}$U is the largest 
collision system at RHIC reaching the highest energy density central 
collisions~\cite{Schenke:2014tga}. In contrast to the nearly or 
completely spherical geometries of the Cu, Au, and Pb 
nuclei~\cite{Adcox:2001jp,Adams:2003kv,Adler:2006bv,Adare:2008qa,Adare:2012uk,Adare:2008ad,Adare:2010dc,Abelev:2012hxa,CMS:2012aa,Aad:2012vca}, 
$^{238}$U is highly deformed. This feature makes U$+$U collisions 
particularly interesting for jet-quenching studies. However, when 
comparing physics observables in U$+$U with Cu$+$Cu, Au$+$Au, or Pb$+$Pb 
collisions, one has to be aware that in any finite collision centrality 
bin the fluctuations of the overlap volume and energy density are larger 
in U$+$U than in the case of spherical nuclei.

\section{Data analysis}

All results presented in this paper were obtained with the PHENIX 
spectrometer from data collected in the Year-2012 data taking period at 
RHIC. A detailed description of the PHENIX experimental set-up can be 
found elsewhere~\cite{Adcox:2003zm}. Event selection is performed with 
two beam-beam counters (BBCs)~\cite{Allen:2003zt} located towards the 
north and south beam directions in the $3.0<|\eta|<3.9$ pseudorapidity 
interval. The collision vertex coordinate along the beam direction 
($z_{{\rm BBC}}$) is determined by the time difference between two hits 
in the north and south BBCs with an accuracy of 0.6--2 cm (depending on 
the particle multiplicity). The analyzed data set was taken with the 
minimum-bias (MB) trigger, which required a north-south coincidence and 
an online vertex position within $\pm30$\,cm. After offline 
reconstruction an additional cut of $|z_{{\rm BBC}}|<20$\,cm was 
applied; the remaining data set comprises $9.4\times10^8$ events.

The event centrality is derived from the distribution of the total 
charge in the BBCs. For each centrality class the mean values of the 
collision geometry parameters, such as the number of binary inelastic 
collisions ($N_{\rm coll}$), participating nucleons ($N_{\rm part}$), 
and $T_{AA}$ (the nuclear overlap integral) are determined using a 
Glauber model based Monte-Carlo simulation of BBC charge 
response~\cite{Miller:2007ri}. For asymmetric $^{238}$U nuclei the 
$\theta$-dependent Woods-Saxon density distribution is used:
\begin{equation}
    \rho(r, \theta)/\rho_0 = \frac{1}{1+\exp{\large[(r-R'(\theta))/a\large]}},
\end{equation}
where $\rho_0$ is the density at the center of the nucleus, $a$ is the 
diffusion parameter, $R'(\theta)=R(1+\beta_2 Y^0_2(\theta)+\beta_4 
Y^0_4(\theta))$, $Y^0_2(\theta)$ and $Y^0_4(\theta)$ are the Legendre 
polynomials. Because there is no single universally accepted 
parametrization of the U$+$U nucleus, we followed the example 
of~\cite{Adare:2015hva,Adamczyk:2015obl}, and used the same two 
parameter sets. Accordingly, two Monte-Carlo simulations were produced 
incorporating different parametrizations of $R'(\theta)$ (see 
Table~\ref{tab:WS}) and, thus, two sets (Glauber~1~\cite{Masui:2009qk} 
and Glauber~2~\cite{Shou:2014eya}) of collision-geometry parameters 
are used, listed in Table~\ref{tab:glauber}.  The obtained \Npart 
values are the same in central collisions and are slightly different in 
more peripheral collisions.  When comparing hadron yields or \raa between 
U$+$U and Au$+$Au collisions at centralities with similar \Npart we have to 
keep in mind, that the rms of the \Npart distribution is wider in U$+$U 
than in Au$+$Au.

\begin{table}[htb]
\caption{\label{tab:WS}
Parameters for the Woods-Saxon distributions used for \uu Glauber 
Monte-Carlo simulations. 
}
\begin{ruledtabular} 
\begin{tabular}{ccc}
Parameter & Glauber~1~\cite{Masui:2009qk} & Glauber~2~\cite{Shou:2014eya} \\
\hline
$R$ (fm) & 6.81 & 6.86 \\
$a$ (fm) & 0.60 & 0.42 \\
$\beta_2$ & 0.280 & 0.265 \\
$\beta_4$& 0.093 & 0 \\
\end{tabular}
\end{ruledtabular}
\end{table}

\begin{table}[htb]
\caption{\label{tab:glauber}
The mean values of $\left\langle T_{AA} \right\rangle$ and the mean
number of participating nucleons $\left\langle N_{\rm part}
\right\rangle$ in different \uu centrality intervals. The values are
shown with their systematic uncertainties, estimated by varying
different input parameters and by using different nucleon density
profiles in the Monte-Carlo Glauber simulations.
}
\begin{ruledtabular} \begin{tabular}{cccc}
Glauber & Centrality interval
& $\left\langle T_{AA} \right\rangle$ (mb$^{-1}$)
& $\left\langle N_{\rm part} \right\rangle$ \\
\hline
Glauber~1~\protect\cite{Masui:2009qk} &
  Minimum Bias & $8.2\pm 1.6$ & $143\pm 5$ \\
&  0\%--20\% & $22.1\pm 2.3$ & $330\pm 6$ \\
& 20\%--40\% & $7.9\pm 0.8$ & $159\pm 7$ \\
& 40\%--60\% & $2.3\pm 0.3$ & $64.8\pm 5.9$ \\
& 60\%--80\% & $0.41\pm 0.09$ & $17.8\pm 3.2$ \\
& 40\%--80\% & $1.34\pm 0.20$ & $41.3\pm 4.5$ \\
Glauber~2~\protect\cite{Shou:2014eya} &
Minimum Bias & $8.9\pm 1.0$ & $144\pm 5$ \\
&  0\%--20\% & $23.7\pm 2.7$ & $330\pm 6$ \\
& 20\%--40\% & $8.9\pm 1.1$ & $161\pm 7$ \\
& 40\%--60\% & $2.6\pm 0.4$ & $65.8\pm 5.8$ \\
& 60\%--80\% & $0.47\pm 0.10$ & $18.2\pm 3.2$ \\
& 40\%--80\% & $1.54\pm 0.22$ & $42.0\pm 4.5$ \\
\end{tabular}
\end{ruledtabular}
\end{table}

\begin{figure*}[tbh]
\includegraphics[width=0.85\linewidth]{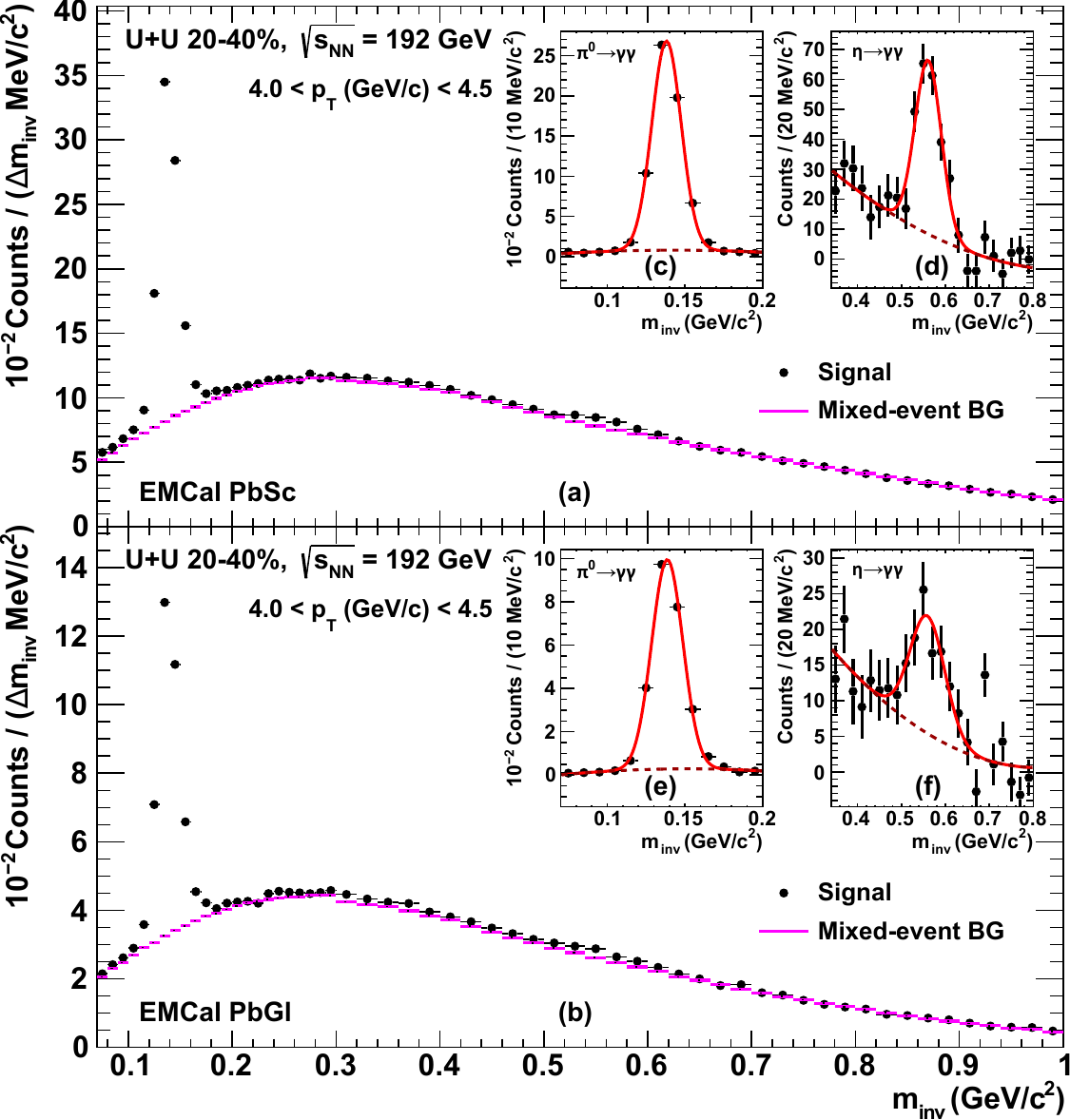}
\caption{Invariant mass distributions for $\gamma\gamma$ pairs, obtained 
in 4--4.5 GeV/$c$ \pt interval in 20--40\% centrality \uu collisions. 
Panels~(a) and (b) show the signal and normalized mixed-event background 
invariant mass distributions in PbSc and PbGl subsystems, respectively. 
In label captions, $\Delta m_{\rm inv}$ stands for the invariant mass 
bin width and is equal to 10 MeV/$c^2$ for $m_{\rm inv} < 0.3$ GeV/$c^2$ 
and for 20 MeV/$c^2$ at larger $m_{\rm inv}$ values. Inserts (c) and (d) 
show the invariant mass distributions in $\pi^0$ and $\eta$ regions 
after the mixed-event background subtraction in PbSc, while inserts (e) 
and (f) show ones in PbGl.
}
\label{fig:Fig1_pi0_eta_peaks}
\end{figure*}

Invariant yields of $\pi^0$ and $\eta$
mesons are obtained from: 
\begin{equation}
   \frac{1}{N_{\rm event}}\frac{d^2N}{2\pi p_T dp_T dy} = 
\frac{N_{\rm raw}}{2\pi p_T N_{\rm event} \epsilon_{\rm rec} 
\Delta p_T \Delta y},
\end{equation}
where $N_{\rm raw}$ is the particle raw yield, $\epsilon_{\rm rec}$ is 
the efficiency (including acceptance and all other corrections), and 
$N_{\rm event}$ is the number of analyzed events.

\begin{table*}[!htb]
\caption{\label{tab:sys_sources}
Sources of systematic uncertainties for $\pi^0$ and $\eta$ 
yields at different \pt. 
Values are shown for 
PbSc(PbGl) subsystems. The types of uncertainties are described in the 
text. Values with a range indicate the variation of the uncertainty over 
the different centrality intervals. 
}
\begin{ruledtabular} \begin{tabular}{ccccc}
Yield & Source             & 2.75 GeV/$c$                      & 13 GeV/$c$                    & Type \\
\hline
$\pi^0\rightarrow\gamma\gamma$
&Acceptance                & 1.5$\%$(1.5$\%$)                  & 1.5$\%$(1.5$\%$)                    & B \\
&$p_T$ weights             & 1$\%$(1$\%$)                      & 1$\%$(1$\%$)                        & B \\
&Energy scale              & 5$\%$(5$\%$)                      & 7$\%$(7$\%$)                        & B \\
&Energy resolution         & 2$\%$(2$\%$)                      & 2$\%$(2$\%$)                        & B \\
&Photon conversion         & 5.2$\%$(5.2$\%$)                  & 5.2$\%$(5.2$\%$)                    & C \\
&Cluster merging           & $-$                               & 7$\%$(4$\%$)                        & B \\
&PID cuts                  & 1.6$\%$(4$\%$)--4$\%$(4$\%$)     & 4$\%$(4$\%$)--6$\%$(4$\%$)         & B \\
&Raw yield extraction      & 1$\%$(1$\%$)--3$\%$(2$\%$)       & 2$\%$(2$\%$)                       & B \\
&Reconstruction efficiency & 0.8$\%$(1.3$\%$)--1.3$\%$(2.0$\%$) & 0.3$\%$(0.4$\%$)--0.4$\%$(0.8$\%$) & A \\

$\eta\rightarrow\gamma\gamma$
&Acceptance                & 1.5$\%$(1.5$\%$)                  & 1.5$\%$(1.5$\%$)                    & B \\
&$p_T$ weights             & 1$\%$(1$\%$)                      & 1$\%$(1$\%$)                        & B \\
&Energy scale              & 3$\%$(3$\%$)                      & 6$\%$(6$\%$)                        & B \\
&Energy resolution         & 2$\%$(2$\%$)                      & 2$\%$(2$\%$)                        & B \\
&Photon conversion         & 5.2$\%$(5.2$\%$)                  & 5.2$\%$(5.2$\%$)                    & C \\
&PID cuts                  & 5$\%$(5$\%$)--5$\%$(7$\%$)       & 5$\%$(5$\%$)                        & B \\
&Raw yield extraction      & 11$\%$(11$\%$)                    & 8$\%$(8$\%$)                        & B \\
&Reconstruction efficiency & 1.2$\%$(2.5$\%$)--3$\%$($5.4\%$)    & 0.4$\%$(0.7$\%$)--0.9$\%$(1.4$\%$) & A \\
\end{tabular} \end{ruledtabular}
\end{table*}

\begin{table*}[!htb]
\caption{\label{tab:sys_total}
Total uncertainties for $\pi^0$ and $\eta$ meson spectra, $R_{AA}$ and 
$\eta/\pi^0$ ratios at different \pt. The types of uncertainties are 
described in the text. Values with a range indicate the variation of the 
uncertainty over the different centrality intervals.
}
\begin{ruledtabular} \begin{tabular}{cccc}
Spectra & Type & $2.75$ GeV/$c$ & $13$ GeV/$c$ \\
\hline
$\pi^0$ PbSc(PbGl) spectra
&stat & 0.3$\%$(0.4$\%$)--0.5$\%$(0.9$\%$) &   6$\%$(8$\%$)--20$\%$(14$\%$)  \\
&A &     0.8$\%$(1.3$\%$)--1.3$\%$(2.0$\%$) & 0.3$\%$(0.4$\%$)--0.4$\%$(0.8$\%$) \\
&B &     6$\%$(7$\%$)--7$\%$(7$\%$) &  12$\%$(10$\%$) \\
&C &          5.2$\%$(5.2$\%$) &  5.2$\%$(5.2$\%$)  \\
$\pi^0$ Combined spectra
&stat   &  0.2\%--0.5\% &   5\%--10\%  \\
&A &  0.7\%--1.1\% & 0.2\%--0.4\%  \\
&B &            6\% &   9\%--10\%  \\
&C &          5.2\% &        5.2\%  \\
$\pi^0$ $R_{AA}$
&A + stat &  0.8\%--1.2\% &   7\%--11\%  \\
&B         &           10\% &  14\%--15\%  \\
&C         &    15\%--26\% &  15\%--26\%  \\
$\eta$ PbSc(PbGl) spectra
&stat & 6$\%$(9$\%$)--8$\%$(14$\%$) &   22$\%$(26$\%$)--32$\%$(36$\%$)  \\
&A &     1.2$\%$(2.0$\%$)--3.0$\%$(5.4$\%$) & 0.4$\%$(0.7$\%$)--0.9$\%$(1.4$\%$) \\
&B &     13$\%$(13$\%$)--13$\%$(14$\%$) &  11$\%$(11$\%$) \\
&C &          5.2$\%$(5.2$\%$) &  5.2$\%$(5.2$\%$)  \\
$\eta$ Combined spectra
&stat   &      5\%--8\%  &   16\%--21\%  \\
&A &    1.2\%--3\%  & 0.3\%--0.7\%  \\
&B &      9\%--10\% &   8\%--9\%  \\
&C &          5.2\%  &        5.2\%  \\
$\eta$ $R_{AA}$
&A + stat &  7\%--10\% &  18\%--23\%  \\
&B         &        19\% &         14\%  \\
           & 15\%--26\% &  15\%--26\%  \\
$\eta/\pi^0$
&A + stat &  5\%--8\%  &  17\%--22\%  \\
&B + C     & 10\%--14\% &         15\%  \\
\end{tabular} \end{ruledtabular}
\end{table*}

\begin{table*}[!htb]
\caption{\label{tab:spectra_fits}
Parameters for the $\pi^0$ and $\eta$ meson invariant 
transverse momentum spectra fits in U$+$U collisions at 
$\sqrt{s_{_{NN}}}=192$ GeV. Only statistical uncertainties are shown.
}
\begin{ruledtabular}
\begin{tabular}{cccccc}
Meson & $p_T$ Limit & Centrality Interval & $A$ & $n$ & $\chi^2/NDF$ \\
\hline
  $\pi^0\rightarrow\gamma\gamma$ & $p_T>5$ GeV/$c$
 & MB & $23.2\pm1.1$ & $8.02\pm0.03$ &  $20.2/12$  \\
&&0\%--20\%       & $44\pm3$     & $7.96\pm0.04$ &  $20.6/11$  \\
&&20\%--40\%      & $38\pm3$     & $8.16\pm0.04$ &  $5.37/11$  \\
&&40\%--60\%      & $13.5\pm1.8$ & $8.06\pm0.07$ &  $1.90/6$    \\
&&60\%--80\%      & $3.7\pm1.0$  & $8.15\pm0.15$ &  $3.82/5$  \\
  $\eta\rightarrow\gamma\gamma$ & $p_T>5$ GeV/$c$
 & MB & $8.1\pm2.6$  & $7.83\pm0.16$ &  $8.19/5$  \\
&&0\%--20\%       & $10\pm6$     & $7.53\pm0.27$ &  $3.65/5$  \\
&&20\%--40\%      & $19\pm10$    & $8.11\pm0.26$ &  $2.89/4$  \\
&&40\%--60\%      & $2.8\pm2.4$  & $7.5\pm0.5$   &  $0.41/1$    \\
\end{tabular} \end{ruledtabular}
\end{table*}

\begin{figure*}[!tbh]
\begin{minipage}[tbh]{0.99\linewidth}
\includegraphics[width=0.9\linewidth]{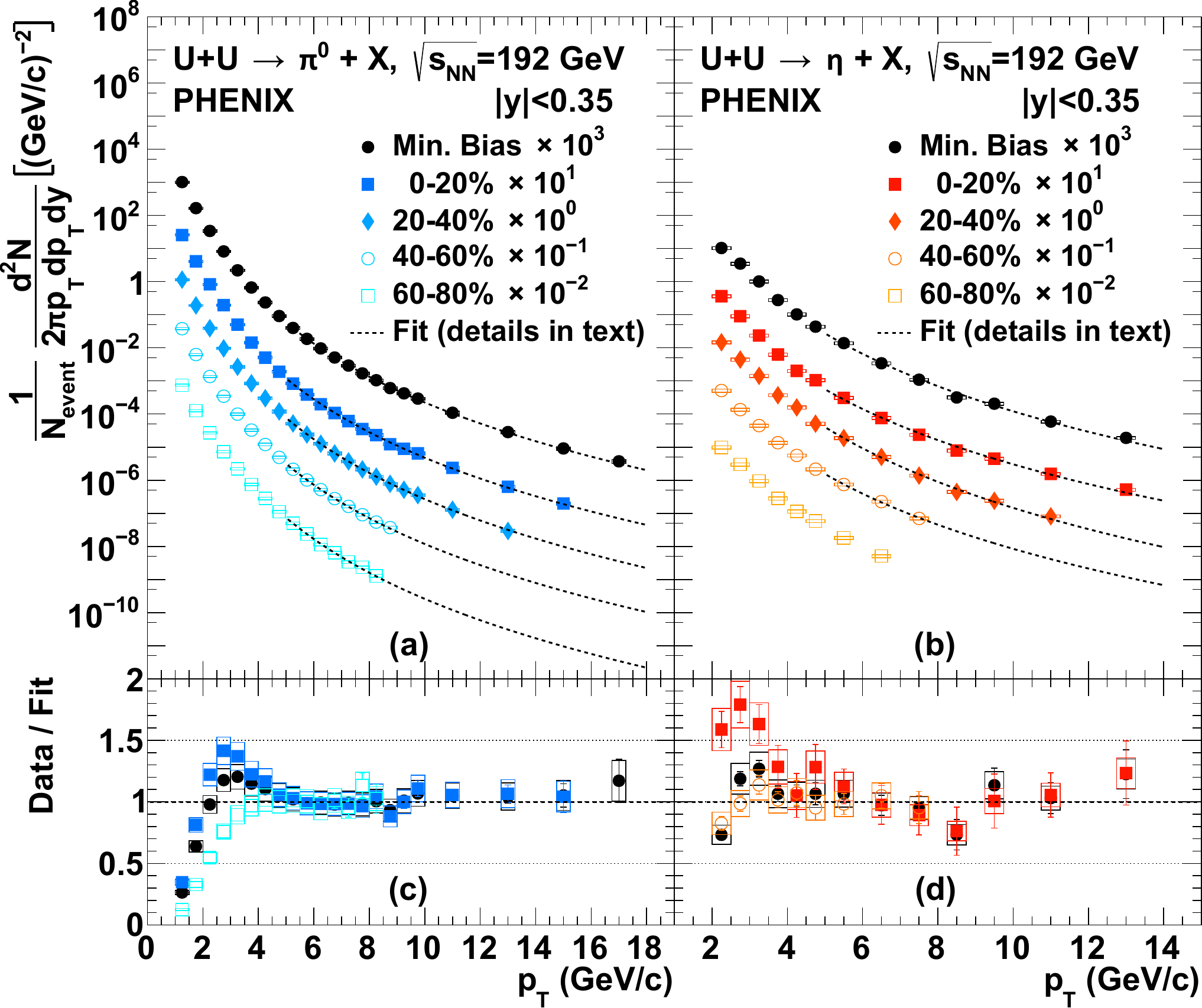}
\caption{\label{fig:Fig2_pi0_eta_invariant_pT_spectra}
$\pi^0$ (a) and $\eta$ (b) invariant $p_T$-spectra 
measured in different centrality intervals of \uu collisions at 
$\sqrt{s_{_{NN}}}=192$ GeV. The dashed curves are fit with a power-law 
function. Error bars represent a quadratic sum of statistical and type-A 
systematic uncertainties. Error boxes represent a quadratic sum of 
type-B and type-C systematic uncertainties. Panels~(c) and (d) 
shows data-to-fit ratios, the markers, error bars and error boxes are, 
respectively, the same as for panels~(a) and (b).
}
\end{minipage}
\begin{minipage}[tbh]{0.99\linewidth}
\vspace{0.5cm}
\includegraphics[width=0.99\linewidth]{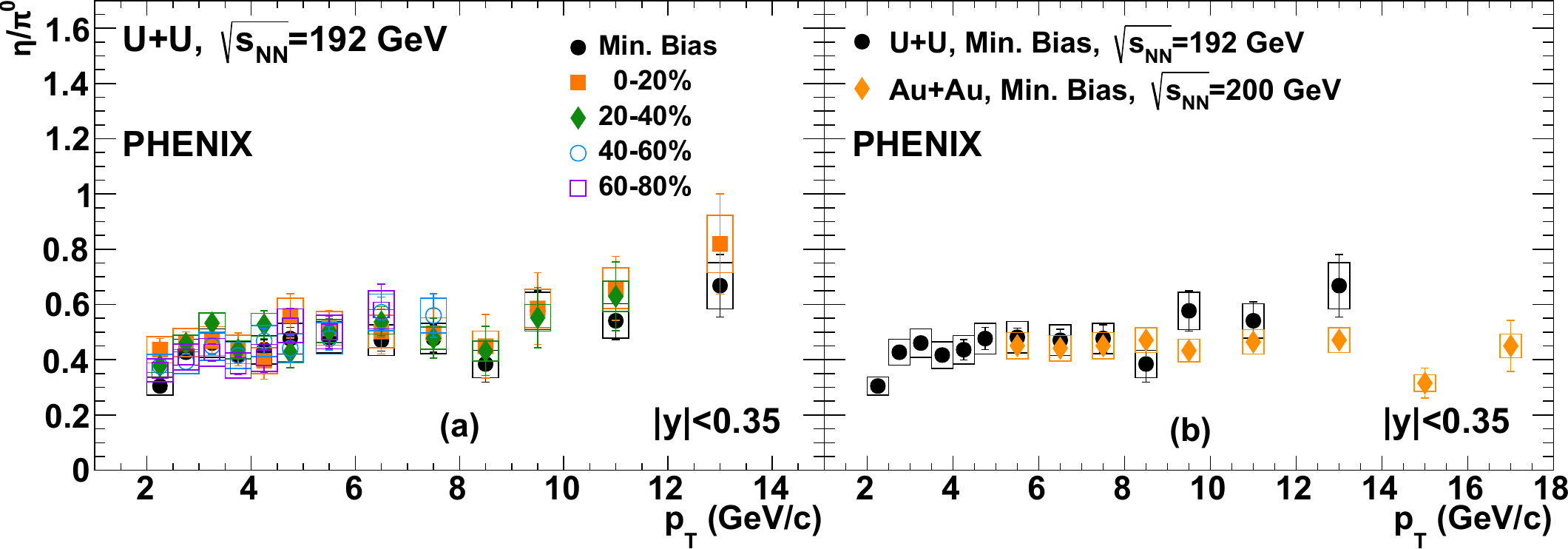}
\caption{\label{fig:Fig3_eta_to_pi0_ratios}
Panel~(a): ratios of $\eta$ and $\pi^0$ yields measured as a 
function of $p_T$ in different centrality intervals of U$+$U 
collisions at $\sqrt{s_{_{NN}}}=192$ GeV. Panel~(b): comparison of 
$\eta$ and $\pi^0$ yields ratios measured as a function of $p_T$ 
in MB U$+$U collisions at $\sqrt{s_{_{NN}}}=192$ GeV and Au$+$Au 
collisions at $\sqrt{s_{_{NN}}}=200$ GeV~\cite{Adare:2012wg} Error 
bars represent a quadratic sum of statistical and type-A 
systematic uncertainties for $\pi^0$ and $\eta$ yields. Error 
boxes represent a quadratic sum of type-B systematic uncertainties 
from $\pi^0$ and $\eta$ yields.
}
\end{minipage}
\end{figure*}

The $\pi^0$ and $\eta$ mesons were reconstructed via the 
$\pi^0\rightarrow\gamma\gamma$ and $\eta\rightarrow\gamma\gamma$ decay 
channels using the electromagnetic calorimeter 
(EMCal)~\cite{Aphecetche:2003zr}. The EMCal comprises two 
technologically different subsystems: lead-scintillator sampling 
calorimeter (PbSc) in four sectors in the west and two sectors in the 
east PHENIX arms, and lead-glass \v{C}erenkov calorimeter (PbGl) in two 
sectors in the east PHENIX arm. Each sector covers $|\eta|<0.35$ 
pseudorapidity range and 22.5 degrees in azimuth. The subsystems have 
different nonlinearity, energy resolution ($\delta 
E/E=2.1\%\oplus8.1\%/\sqrt{E}$ for PbSc and $0.8\%\oplus5.9\%/\sqrt{E}$ 
for PbGl) and segmentation ($\delta\phi\times\delta\eta \approx 
0.01\times0.01 $ for PbSc and $0.008\times0.008 $ for PbGl).

Showers in the EMCal are selected as $\gamma$ candidates if they pass a 
shower shape cut~\cite{Aphecetche:2003zr} and a minimum energy cut 
($E_{\gamma\min}=0.4$ GeV) to reduce contamination from minimum ionizing 
hadrons. Then $\gamma\gamma$ pairs are formed from all photon candidates 
in the same sector under the condition that their energies ($E_{\gamma 
1}$ and $E_{\gamma 2}$) satisfy an asymmetry cut $|E_{\gamma 
1}-E_{\gamma 2}|/(E_{\gamma 1}+E_{\gamma 2})<0.8$, to reduce 
the combinatorial background.

To determine raw yields of $\pi^0$ and $\eta$ mesons the invariant mass 
($m_{\rm inv}$) distributions of $\gamma\gamma$ pairs passing the cuts are 
produced in different $p_T$ and centrality intervals, separately for 
PbSc and PbGl subsystems~\cite{Adcox:2001jp}. The distributions contain 
a background and two signal peaks around $m_{\rm inv}\approx 0.14$ and 
$0.55$ GeV/$c^2$, corresponding to $\pi^0$ and $\eta$ decays, 
respectively. The background comprises correlated and uncorrelated 
components. The correlated component comes from photons of other 
particle decays ($K_S$, $\omega$, $\rho$, $\eta '$ etc). The 
uncorrelated component of the background comes from combinations of 
uncorrelated $\gamma$ candidates and is well reproduced by event mixing, 
where $\gamma\gamma$ pairs are formed from two $\gamma$ candidates from 
different events with similar collision vertex ($z_{{\rm BBC}}$) and 
centrality. Estimated background shapes are normalized to the real 
(same-event) $\gamma\gamma$ $m_{\rm inv}$ distributions in the ranges 
$0.08<m_{\rm inv}<0.085$ and $0.36<m_{\rm inv}<0.40$ GeV/$c^2$ for the 
$\pi^0$, in $0.7<m_{\rm inv}<0.8$ GeV/$c^2$ for the $\eta$, and then 
subtracted. Due to the rapid decrease of the combinatorial background 
with increasing $p_T$, the mixed-event subtraction is implemented only 
for $p_T<10$ GeV/$c$. Typical $\gamma\gamma$ invariant mass 
distributions and corresponding normalized mixed-event background shapes 
are presented in Fig.~\ref{fig:Fig1_pi0_eta_peaks}, where panels~(a) and 
(b) correspond to PbSc and PbGl measurements, respectively. Note, that 
in Fig.~\ref{fig:Fig1_pi0_eta_peaks} the foreground and background 
distributions plotted at $m_{\rm inv} < 0.3$ GeV/$c^2$ correspond to 
$\pi^0$ meson measurement, while at higher $m_{\rm inv}$ values ones 
correspond to $\eta$ measurements, so the distributions have different 
binwidth in the two invariant mass ranges.

The resulting $m_{\rm inv}$ distributions are fitted to a 
combination of a Gaussian and a polynomial to describe a signal 
and the residual (correlated) background, respectively. For 
$\pi^0$ and $\eta$ measurements, respectively, first and second 
order polynomials were used. Meson raw yields are determined as 
the difference between the integrals of the bin content and the 
polynomial in the mass peak regions, which are defined as 
$0.10<m_{\rm inv}<0.17$ and $0.48<m_{\rm inv}<0.62$ GeV/$c^2$ for 
$\pi^0$ and $\eta$ peaks, respectively. Panels~(c), (d), (e), and 
(f) of Fig.~\ref{fig:Fig1_pi0_eta_peaks} present examples of the 
resulting $m_{\rm inv}$ distributions in the $\pi^0$ (panels~(c) 
and (e)) and $\eta$ (panels~(e) and (f)) regions obtained in PbSc 
(panels~(c) and (d)) or PbGl (panels~(e) and (f)) subsystems as well as 
the corresponding fitting functions examples.

Acceptance and reconstruction efficiency (efficiency hereafter) is 
estimated using a {\sc geant3}-based~\cite{Brun:1978fy} Monte-Carlo 
simulation of the PHENIX detector. The simulation is tuned to reproduce 
the observed mass peaks and widths of $\pi^0$ and $\eta$ mesons
in the real data. To account for the effect of underlying events 
(multiplicity) the simulated mesons are embedded in real data in each 
centrality, then analyzed with the same methods as the real data. Final 
efficiencies also account for branching ratios of the meson decay modes.

Systematic uncertainties of the measurements are classified into three 
types. Type-A uncertainties are entirely \pt-uncorrelated and are added 
in quadrature to the statistical uncertainties. Type-B uncertainties are 
\pt-correlated, but different from point to point, and all data points 
can move up or down by the same fraction of their type-B uncertainty. 
Type-C uncertainties move all points up or down by the same 
fraction~\cite{Adare:2008cg}.

Sources of systematic uncertainties for $\pi^0$ and $\eta$ yield
measurements are listed in Table~\ref{tab:sys_sources} for 
representative \pt values. Examples of total uncertainties of different 
types for the meson spectra, \raa, and ratios are listed 
in Table~\ref{tab:sys_total}.

In $\pi^0$ measurements, the main sources of systematic uncertainty at 
low \pt (1--3 GeV/$c$) are photon conversions in the detector 
material, at intermediate \pt (3--12 GeV/$c$) the absolute energy 
calibration of the EMCal, and at high \pt ($>$12 GeV/$c$) the cluster 
merging effect. The uncertainty on the absolute scale comes from the 
approximately $1\%$ residual mismatch between $\pi^0$ masses in real 
data and simulation. This causes a systematic uncertainty that increases 
gradually from $2\%$ at low \pt, $7\%$ at intermediate \pt and $9\%$ at 
the highest momenta. Cluster merging is due to the small opening angle 
of daughter photons of the high-$p_T$ $\pi^0$, so these photons are 
reconstructed as a single electromagnetic cluster and the \piz is lost. 
The cluster merging effect starts at $p_T>12$ GeV/$c$ in PbSc and at 
$p_T>16$ GeV/$c$ in PbGl and results in uncertainty reaching 
$\approx$20$\%$ and $\approx$9$\%$ at 20 GeV/$c$ for $\pi^0$ yields, 
reconstructed in PbSc and PbGl subsystems, respectively. For $\eta$ 
mesons, which have a four times larger mass than $\pi^0$, the cluster 
merging effect would be significant starting at 50 GeV/$c$, which is far 
beyond the $p_T$ range of $\eta$ measurement at PHENIX. 

For $\eta$ measurements the dominant systematic uncertainty comes from 
the raw yield extraction. The uncertainty is connected to the selection 
of the invariant mass distributions analysis parameters such as the 
fitting range, the background normalization, the polynomial order 
selection etc. The maximum difference between the meson yield obtained 
with the varied parameters and the one obtained with the default 
parameters is assigned as an uncertainty on raw yield extraction, and it 
varies from 7\% to 12\% for the $\eta$ yields depending on $p_T$ and 
centrality (see Table~\ref{tab:sys_sources}).

Systematic uncertainties for $\eta/\pi^0$ ratios are calculated as a 
quadratic sum of the type-B uncertainties from $\pi^0$ and $\eta$ 
yields. Because type-C uncertainties of the $\pi^0$ and $\eta$ yields 
are 100\% correlated between these particle measurements for all \pt, 
this uncertainty cancels in the ratios. The $p_T$-correlated systematic 
uncertainties for $R_{AA}$ include both uncertainties from U$+$U and 
$p$+$p$ 
measurements~\cite{Adare:2008qb,Adare:2007dg,Adare:2015ozj,Adler:2006bv}. 
Examples of total uncertainties of different types for the meson 
spectra, \raa, and ratios are listed in 
Table~\ref{tab:sys_total}.

In $\pi^0$ and $\eta$ measurements, the presented invariant yields are 
obtained by averaging the PbSc and PbGl results. The averaging uses 
weights defined by the quadratic sum of statistical and uncorrelated 
systematic uncertainties. Please note that uncertainties, which are 
correlated between two subsystems (like conversion) were added after the 
averaging. A comparison between the PbSc(PbGl) spectra uncertainties and 
the combined ones are shown in Table~\ref{tab:sys_total}. Data points 
are plotted at the bin centers rather than the bin-averaged position to 
facilitate a comparison between different experiments and data sets. To 
represent the true physical values at the $p_T$ of the bin center, the 
data have been adjusted to correct for nonlinear effects in 
bin-averaging on a steeply falling spectrum~\cite{Lafferty:1994cj}.

\section{Results and discussion}

Invariant $p_T$ spectra for $\pi^0$ and $\eta$ mesons in different 
U$+$U collision centrality intervals and MB collisions are shown 
in panels~(a) and (b) of 
Fig.~\ref{fig:Fig2_pi0_eta_invariant_pT_spectra}, respectively. 
At low \pt the measurements are limited by the rapidly decreasing 
$S/B$ ratio, and at high \pt by the available statistics. In 
central U$+$U collisions $\pi^0$ and $\eta$ yields are measured up 
to 16 and 14 GeV/$c$, respectively. At $p_T>5$ GeV/$c$ the meson 
spectra are fitted to the power-law function:

\begin{equation}
f(p_T) = \frac{A}{p_T^n},
\end{equation}
where $A$ and $n$ are free parameters. The estimated values of 
these parameters and the $\chi^2/NDF$ values are listed in 
Table~\ref{tab:spectra_fits} for each meson species and centrality 
interval of U$+$U collisions

The $\eta/\pi^0$ ratios ($R_{\eta/\pi^0}$) as a function of $p_T$ 
for different U$+$U centrality intervals are presented in 
panel~(a) of Fig.~\ref{fig:Fig3_eta_to_pi0_ratios}. The comparison 
of $\eta/\pi^0$ ratios obtained in U$+$U and 
Au$+$Au~\cite{Adare:2012wg} collisions is shown in panel~(b) of 
the same figure.  Within uncertainties the measured 
$R_{\eta/\pi^0}$ are independent of centrality in the whole $p_T$ 
range. A constant fit to the MB data at $p_T > 4$ GeV/$c$ for 
$\eta/\pi^0$ results in $\eta/\pi^0 = 0.476 \pm 0.016$, and the 
various centrality bins are consistent with this value. Similar 
results were obtained in hadron-hadron, hadron-nucleus, 
nucleus-nucleus and $e^{+}e^{-}$ collisions in a wide range of 
collision energies $\sqrt{s_{_{NN}}}=3$--2760 GeV (see for 
instance~\cite{Busser:1974yj,Apanasevich:2002wt,Adler:2006bv,Aidala:2018ond,Acharya:2018yhg}). 
This suggests that the QGP medium produced in U$+$U collisions 
either does not affect the jet fragmentation into light mesons (it 
is similar as in vacuum) or it affects the \piz and $\eta$ the 
same way, despite their different flavor content.

\begin{figure*}[ht]
\begin{minipage}[tbh]{0.99\linewidth}
\includegraphics[width=0.8\linewidth]{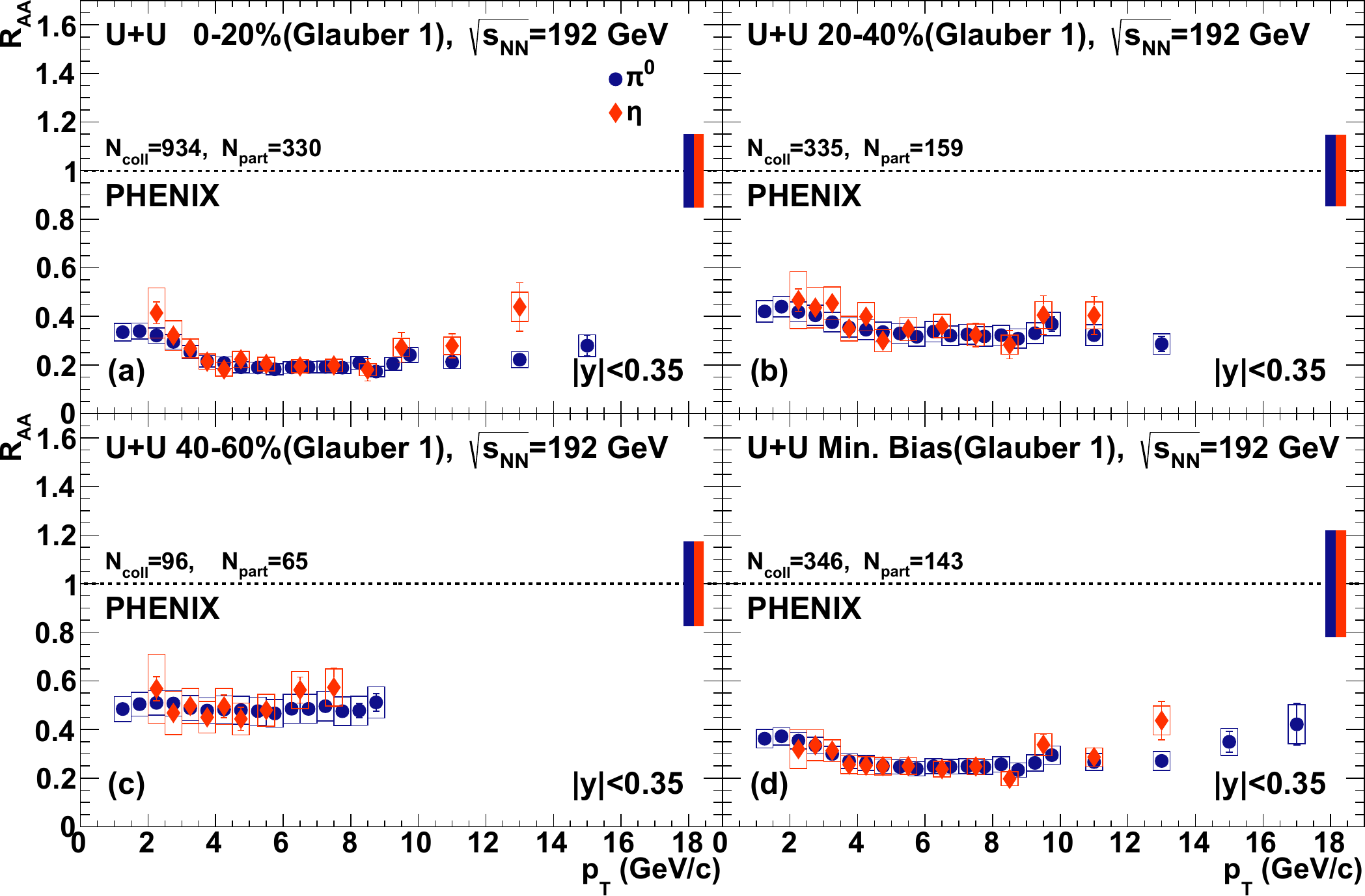}
\caption{ $R_{AA}$ of $\pi^0$ and $\eta$ mesons measured as a 
function of $p_T$ in different centrality intervals of U$+$U collisions 
at $\sqrt{s_{_{NN}}}=192$ GeV. Error bars represent a quadratic sum of 
statistical and type-A systematic uncertainties from U$+$U and $p$+$p$ 
measurements, respectively. Error boxes represent type-B systematic 
uncertainties from U$+$U and $p$+$p$ measurements. Solid and open boxes 
at unity represent type-C systematic uncertainties from U$+$U (including 
uncertainties from the $T_{AA}$ values) and $p$+$p$, respectively.
}   
\label{fig:Fig4_UU_RAA}
\end{minipage}
\begin{minipage}[tbh]{0.48\linewidth}
\vspace{0.2cm}
\includegraphics[width=0.85\linewidth]{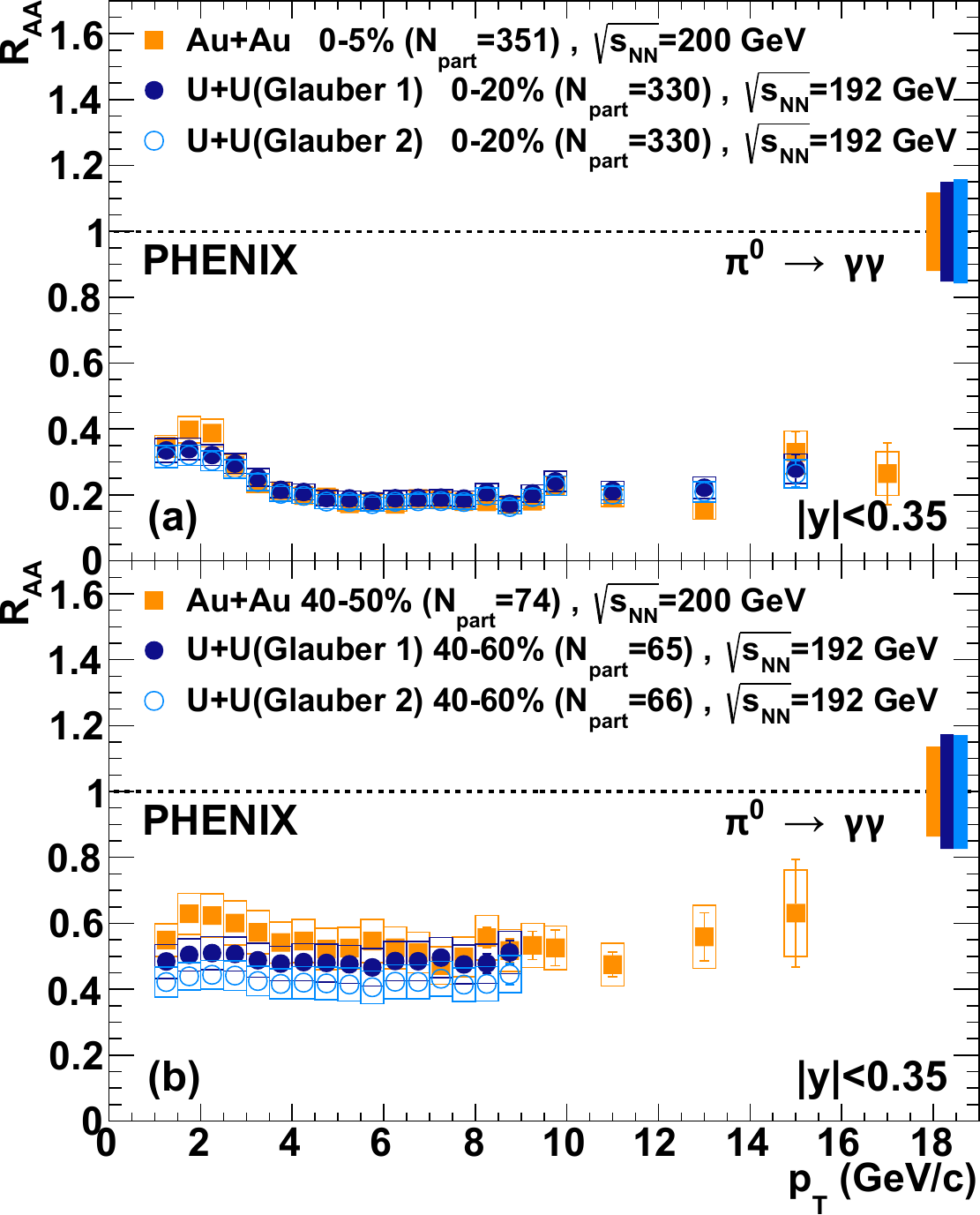}
\caption{Comparison at comparable \Npart of $\pi^0$ $R_{AA}$ 
measured in \uu (two Glauber sets) at $\sqrt{s_{_{NN}}}=192$ GeV 
and in $\auau$ collisions at 
$\sqrt{s_{_{NN}}}=200$ GeV~\cite{Adare:2008qa}.  Error bars 
represent a quadratic sum of statistical and type-A systematic 
uncertainties from U$+$U and $p$$+$$p$ measurements. Open boxes are 
type-B systematic uncertainties for U$+$U and \pp collisions. The 
three boxes at unity are type-C systematic uncertainties from 
$p$$+$$p$ and nucleus-nucleus collisions. The boxes from left to 
right correspond to Au$+$Au and U$+$U measurements, respectively.
Note that while the mean \Npart is similar for 40--60\% U$+$U and 
40--50\% \auau, its rms is 23.2 for U$+$U while only 13.1 for \auau.
}
\label{fig:Fig5_pi0_RAA_Comparison}
\end{minipage}
\hspace{0.2cm}
\begin{minipage}[tbh]{0.48\linewidth}
\vspace{-0.6cm}
\includegraphics[width=0.85\linewidth]{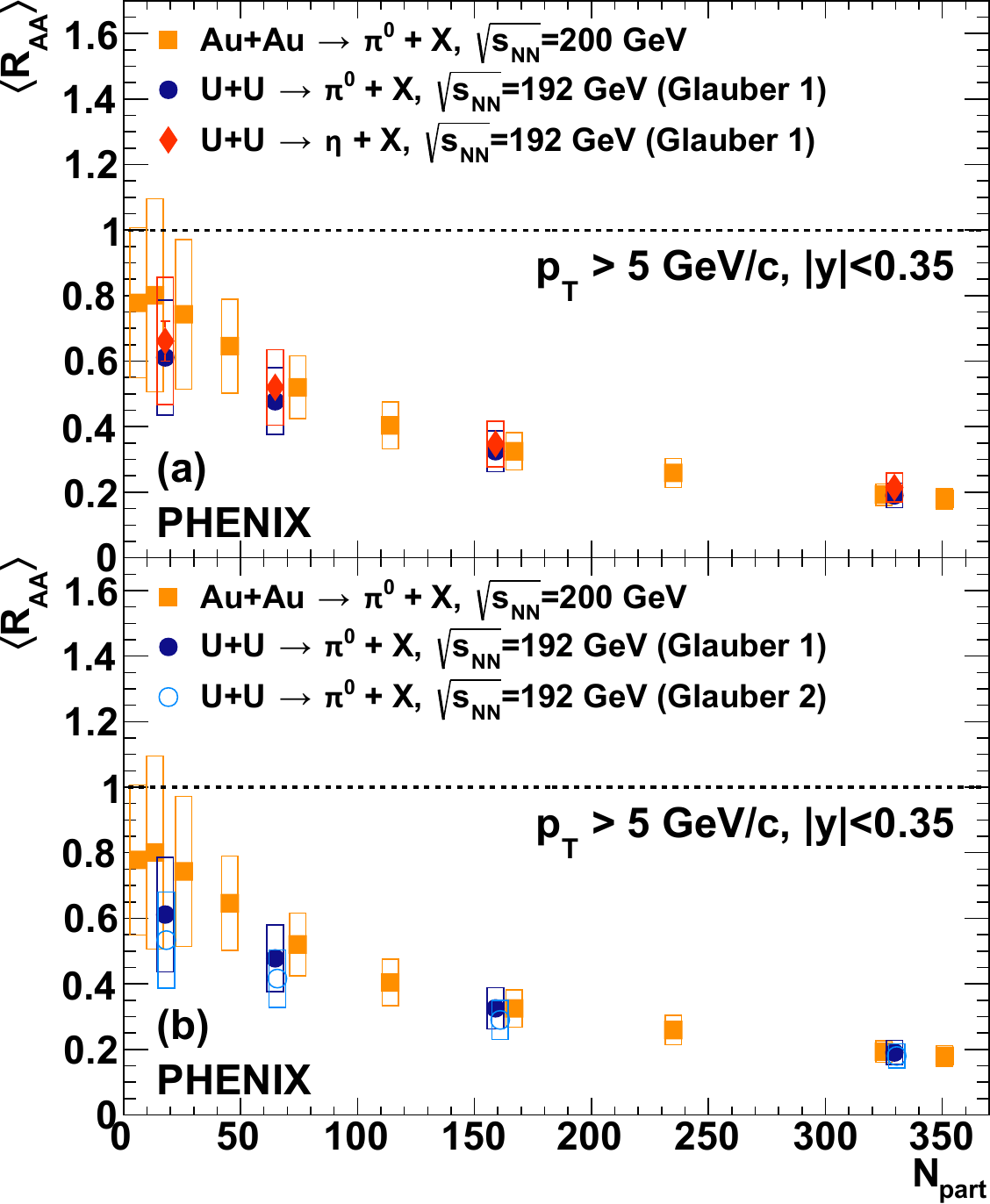}
\vspace{0.25cm}
\caption{Panel~(a): comparison of integrated $R_{AA}$ for \piz and 
$\eta$ measured as a function of $N_{\rm part}$ in \uu (Glauber-1 
set) and $\auau$ collisions at $\sqrt{s_{_{NN}}}=192$ GeV and 
$\sqrt{s_{_{NN}}}=200$ GeV, respectively. Panel~(b): comparison of 
integrated $R_{AA}$ for \piz measured as a function of $N_{\rm 
part}$ in \uu (Glauber-1 and Glauber-2 sets) and $\auau$ 
collisions. Uncertainties are the same as in the 
Fig.~\ref{fig:Fig5_pi0_RAA_Comparison}.  The lower limit of 
integration is $\pt = 5$\,\gevc.
}
\label{fig:Fig7_Int_RAA}
\end{minipage}
\end{figure*}

Figure \ref{fig:Fig4_UU_RAA} shows the \raa of $\pi^0$ and $\eta$ mesons as 
functions of $p_T$ for different 
U$+$U centrality intervals. Results are presented only for the Glauber-1 set, 
the use of the Glauber-2 set will not change the comparison 
between different meson species. To calculate \raa
one needs to use the $p$+$p$ differential cross sections 
obtained at the same energy as the $A$$+$$A$ yields. RHIC does not have 
$p$+$p$ data at $\sqrt{s}=192$ GeV, thus the meson cross sections at 
this energy are estimated assuming their power-law dependence on 
$\sqrt{s}$, using results at available $\sqrt{s}$ values, as it was done 
for charged particles at $\sqrt{s}=5.02$ TeV in 
ALICE~\cite{ALICE:2012mj}. For $\pi^0$ measurement the interpolation is 
carried out from the $p$$+$$p$ data at $\sqrt{s}=62.4$, 200, and 
510~GeV. Table~\ref{tab:cs_pp192} shows the results of the recalculation.

For $\eta$ measurements there are no $p$+$p$ data available at 
$\sqrt{s}=62.4$ and $510$ GeV, thus the cross sections for these mesons 
are recalculated from ones at $\sqrt{s}=200$ GeV~\cite{Adler:2006bv} 
using the ratio between $\pi^0$ cross sections at $\sqrt{s}=192$ and 
$200$ GeV.  The obtained $\pi^0$ and $\eta$ meson \raa
are consistent within uncertainties in the whole $p_T$ range for 
every analyzed centrality interval of U$+$U collisions. At $p_T>5$ 
GeV/$c$ $R_{AA}$ is $\approx 0.2 - 0.3$ in the most central collisions. 
A weak $p_T$ dependence of the measured $R_{AA}$ values can be observed. 
The suppression of $\pi^0$ and $\eta$ mesons decreases as one 
moves to more peripheral collisions.

Figure~\ref{fig:Fig5_pi0_RAA_Comparison} compares \raa of $\pi^0$ 
mesons measured as a function of $p_T$ in \sqsn=192\,GeV \uu for two 
Glauber sets and \sqsn=200\,GeV $\auau$~\cite{Adare:2008qa} collisions, 
plotted for similar $N_{\rm part}$ values. It follows from the \Ncoll 
values listed in Table~\ref{tab:WS} that in peripheral collisions the 
central values of \raa are slightly different for the Glauber-1 and 
Glauber-2 models, however, the difference is within experimental 
uncertainties. The observed $\pi^0$ $R_{AA}$ is the same for U$+$U and 
Au$+$Au collisions within uncertainties, which suggests that the $\pi^0$ 
suppression mostly depends on the energy density and size of the 
produced medium. Note that while the mean \Npart is similar for 40--60\% 
U$+$U and 40--50\% \auau, its rms is 23.2 for U$+$U while only 13.1 for 
\auau.

Figure~\ref{fig:Fig7_Int_RAA} shows the $\pi^0$ and $\eta$ integrated 
\raa as a function of $N_{\rm part}$ for \uu compared to \auau.  
Panel~(b) of Fig.~\ref{fig:Fig7_Int_RAA} presents the comparison of \piz 
integrated \raa as an $N_{\rm part}$ function between $\auau$ and two 
Glauber sets of \uu. The integration is carried out for $\pt>5$\,\gevc. 
Values of obtained integrated $R_{AA}$ are shown in 
Table~\ref{tab:intRAA} for different meson species and for Glauber-1 
set. The results obtained for the two different collision systems are on 
a universal trend as a function of $\Npart$.  The dominant factor in 
this observable is the size of the overlap volume (\Npart), while the 
much larger fluctuations in U$+$U because of its shape are secondary.


\begin{table*}[!htb]
\begin{minipage}[tbh]{0.99\linewidth}
  \caption{\label{tab:cs_pp192}
Production cross section of $\pi^0$ and $\eta$ mesons in $p$+$p$
collisions, recalculated at $\sqrt{s}=192$ GeV.
}
\begin{ruledtabular}   \begin{tabular}{cccccc}
Meson & $p_T$     &  $E\,d^3\sigma/d^3p$  &  Stat + Type-A & Type-B      & Type-C     \\
Decay & (GeV/$c$) &  (mb/GeV$^{-2}c^3$)   &  Uncertainty    & Uncertainty & Uncertainty \\
 \hline
 $\pi^0\rightarrow\gamma\gamma$
&1.25 &   $3.85\times10^{-1}$ & $2.8\times10^{-4}$ &   $3.76\times10^{-2}$ &  $3.74\times10^{-2}$ \\
&1.75 &   $5.97\times10^{-2}$ &   $7\times10^{-5}$ &   $4.92\times10^{-3}$ &  $5.79\times10^{-3}$ \\
&2.25 &   $1.25\times10^{-2}$ & $2.5\times10^{-5}$ &   $1.03\times10^{-3}$ &  $1.22\times10^{-3}$ \\
&2.75 &   $3.16\times10^{-3}$ & $1.0\times10^{-5}$ &   $2.61\times10^{-4}$ &  $3.06\times10^{-4}$ \\
&3.25 &   $9.35\times10^{-4}$ &   $5\times10^{-6}$ &    $7.8\times10^{-5}$ &   $9.1\times10^{-5}$ \\
&3.75 &   $3.12\times10^{-4}$ & $2.5\times10^{-6}$ &   $2.65\times10^{-5}$ &  $3.02\times10^{-5}$ \\
&4.25 &   $1.12\times10^{-4}$ & $2.4\times10^{-7}$ &   $1.03\times10^{-5}$ &  $1.09\times10^{-5}$ \\
&4.75 &   $4.60\times10^{-5}$ & $1.4\times10^{-7}$ &   $4.24\times10^{-6}$ &  $4.46\times10^{-6}$ \\
&5.25 &   $2.02\times10^{-5}$ &   $8\times10^{-8}$ &   $1.88\times10^{-6}$ &  $1.96\times10^{-6}$ \\
&5.75 &   $9.73\times10^{-6}$ &   $6\times10^{-8}$ &    $9.1\times10^{-7}$ &   $9.4\times10^{-7}$ \\
&6.25 &   $4.83\times10^{-6}$ & $3.5\times10^{-8}$ &   $4.52\times10^{-7}$ &  $4.68\times10^{-7}$ \\
&6.75 &   $2.55\times10^{-6}$ & $2.5\times10^{-8}$ &   $2.40\times10^{-7}$ &  $2.47\times10^{-7}$ \\
&7.25 &   $1.44\times10^{-6}$ & $1.8\times10^{-8}$ &   $1.37\times10^{-7}$ &  $1.40\times10^{-7}$ \\
&7.75 &   $8.43\times10^{-7}$ & $1.3\times10^{-8}$ &    $8.0\times10^{-8}$ &   $8.2\times10^{-8}$ \\
&8.25 &   $5.02\times10^{-7}$ & $1.0\times10^{-8}$ &    $4.8\times10^{-8}$ &   $4.9\times10^{-8}$ \\
&8.75 &   $3.19\times10^{-7}$ &   $7\times10^{-9}$ &    $3.1\times10^{-8}$ &   $3.1\times10^{-8}$ \\
&9.25 &   $1.96\times10^{-7}$ &   $6\times10^{-9}$ &    $1.9\times10^{-8}$ &   $1.9\times10^{-8}$ \\
&9.75 &   $1.21\times10^{-7}$ &   $4\times10^{-9}$ &    $1.2\times10^{-8}$ &   $1.2\times10^{-8}$ \\
&11 &     $5.41\times10^{-8}$ & $1.4\times10^{-9}$ &    $5.5\times10^{-9}$ &   $5.2\times10^{-9}$ \\
&13 &     $1.35\times10^{-8}$ &  $6\times10^{-10}$ &    $1.5\times10^{-9}$ &   $1.3\times10^{-9}$ \\
&15 &     $3.31\times10^{-9}$ & $2.8\times10^{-10}$ &   $4.0\times10^{-10}$ &  $3.2\times10^{-10}$ \\
&17 &     $1.11\times10^{-9}$ & $1.5\times10^{-10}$ &   $1.5\times10^{-10}$ &  $1.1\times10^{-10}$ \\
&19 &     $4.8\times10^{-10}$ & $1.1\times10^{-10}$ &     $8\times10^{-11}$ &    $5\times10^{-11}$ \\
 $\eta\rightarrow\gamma\gamma$
&2.25 &   $3.98\times10^{-3}$ & $2.2\times10^{-4}$ &  $9.2\times10^{-4}$ &  $3.9\times10^{-4}$ \\
&2.75 &   $1.28\times10^{-3}$ &   $7\times10^{-5}$ &  $2.1\times10^{-4}$ &  $1.2\times10^{-4}$ \\
&3.25 &   $3.96\times10^{-4}$ & $1.7\times10^{-6}$ & $4.41\times10^{-5}$ & $3.84\times10^{-5}$ \\
&3.75 &   $1.33\times10^{-4}$ &   $8\times10^{-7}$ & $1.50\times10^{-5}$ & $1.29\times10^{-5}$ \\
&4.25 &   $4.99\times10^{-5}$ & $3.8\times10^{-7}$ & $5.73\times10^{-6}$ & $4.84\times10^{-6}$ \\
&4.75 &   $2.14\times10^{-5}$ & $2.1\times10^{-7}$ & $2.47\times10^{-6}$ & $2.08\times10^{-6}$ \\
&5.5 &    $6.80\times10^{-6}$ &   $5\times10^{-8}$ &  $7.0\times10^{-7}$ &  $6.6\times10^{-7}$ \\
&6.5 &    $1.76\times10^{-6}$ & $2.2\times10^{-8}$ & $1.84\times10^{-7}$ & $1.71\times10^{-7}$ \\
&7.5 &    $5.37\times10^{-7}$ & $1.1\times10^{-8}$ &  $5.6\times10^{-8}$ &  $5.2\times10^{-8}$ \\
&8.5 &    $1.96\times10^{-7}$ &   $6\times10^{-9}$ &  $2.1\times10^{-8}$ &  $1.9\times10^{-8}$ \\
&9.5 &    $7.42\times10^{-8}$ & $3.2\times10^{-9}$ &  $7.8\times10^{-9}$ &  $7.2\times10^{-9}$ \\
&11 &     $2.52\times10^{-8}$ & $1.1\times10^{-9}$ &  $2.7\times10^{-9}$ &  $2.4\times10^{-9}$ \\
&13 &     $5.32\times10^{-9}$ & $4.4\times10^{-10}$ & $5.7\times10^{-10}$ & $5.2\times10^{-10}$ \\
&15 &     $1.66\times10^{-9}$ & $2.4\times10^{-10}$ & $1.8\times10^{-10}$ & $1.6\times10^{-10}$ \\
&17 &     $5.5\times10^{-10}$ & $1.2\times10^{-10}$ &   $6\times10^{-11}$ &   $5\times10^{-11}$ \\
   \end{tabular}   \end{ruledtabular}
\end{minipage}
\begin{minipage}[tbh]{0.99\linewidth}
  \caption{\label{tab:intRAA}
Integrated $R_{AA}$ for $\pi^0$ and $\eta$ mesons as a function of 
$N_{\rm part}$ in U$+$U collisions at $\sqrt{s_{_{NN}}}=192$ GeV, 
calculated for Glauber~1.
}
\begin{ruledtabular}
   \begin{tabular}{cccccc}
\multirow{ 2}{*}{Meson} & \multirow{ 2}{*}{$p_T$ Limit} &
\multirow{ 2}{*}{$N_{\rm part}$} &  \multirow{ 2}{*}{$\langle R_{AA} \rangle$}
                        &  Stat + Type-A & Type-B+Type-C \\
 & & & &  Uncertainty & Uncertainty \\
\hline
  $\pi^0\rightarrow\gamma\gamma$ & $p_T > 5$ GeV/$c$
 &$17.8$ & $0.61$ & $0.011$ & $0.18$ \\
&&$64.8$ & $0.48$ & $0.005$ & $0.10$ \\
&&$159$  & $0.33$ & $0.0030$ & $0.063$ \\
&&$330$  & $0.19$ & $0.0016$ & $0.037$ \\
  $\eta\rightarrow\gamma\gamma$ & $p_T > 5$ GeV/$c$
 &$17.8$ & $0.66$ & $0.06$ & $0.19$ \\
&&$64.8$ & $0.52$ & $0.030$ & $0.12$ \\
&&$159$ & $0.35$ & $0.019$ & $0.07$ \\
&&$320$ & $0.22$ & $0.015$ & $0.04$ \\
    \end{tabular} \end{ruledtabular}
\end{minipage}
\end{table*}


\clearpage

\section{Summary}

PHENIX has measured $\pi^0$ and $\eta$ invariant $p_T$ spectra and \raa 
in the heaviest collision system available at RHIC, U$+$U at 
$\sqrt{s_{_{NN}}}=192$ GeV in a wide $p_T$ range ($1 < p_T < 18$ and $2 
< p_T< 14$, respectively) and for several centrality intervals. In the 
more central collisions the spectra are similar to those observed in 
\auau at similar \Npart (the powers $n$ in U$+$U are consistent within 
fitting errors with the respective fitted powers to Au$+$Au 
in~\cite{Adare:2008qa}). The values of $\eta/\piz$ are independent of 
collision centrality and $p_T$, and consistent with the previously 
measured values in hadron-hadron, hadron-nucleus, nucleus-nucleus as 
well as $e^{+}e^{-}$ collisions at $\sqrt{s_{_{NN}}}=$3--2760 GeV, 
suggesting that either the fragmentation of jets into \piz and $\eta$ is 
unchanged, irrespective of the absence or presence of the medium, or it 
changes the same way, despite the different flavor content. The values 
of $R_{AA}$ for \piz and $\eta$ are consistent within uncertainties in 
all analyzed centrality intervals of \uu collisions. The suppression 
pattern of $\pi^0$ in \uu collisions is consistent with $\auau$ 
collisions at the similar interaction energy and similar values of 
$N_{\rm part}$, except for $N_{\rm part} < 100$ and $p_T < 4$ GeV/$c$ 
(see Fig.~\ref{fig:Fig5_pi0_RAA_Comparison}).

\begin{acknowledgments}


We thank the staff of the Collider-Accelerator and Physics
Departments at Brookhaven National Laboratory and the staff of
the other PHENIX participating institutions for their vital
contributions.  We acknowledge support from the 
Office of Nuclear Physics in the
Office of Science of the Department of Energy,
the National Science Foundation, 
Abilene Christian University Research Council, 
Research Foundation of SUNY, and
Dean of the College of Arts and Sciences, Vanderbilt University 
(U.S.A),
Ministry of Education, Culture, Sports, Science, and Technology
and the Japan Society for the Promotion of Science (Japan),
Conselho Nacional de Desenvolvimento Cient\'{\i}fico e
Tecnol{\'o}gico and Funda\c c{\~a}o de Amparo {\`a} Pesquisa do
Estado de S{\~a}o Paulo (Brazil),
Natural Science Foundation of China (People's Republic of China),
Croatian Science Foundation and
Ministry of Science and Education (Croatia),
Ministry of Education, Youth and Sports (Czech Republic),
Centre National de la Recherche Scientifique, Commissariat
{\`a} l'{\'E}nergie Atomique, and Institut National de Physique
Nucl{\'e}aire et de Physique des Particules (France),
Bundesministerium f\"ur Bildung und Forschung, Deutscher Akademischer 
Austausch Dienst, and Alexander von Humboldt Stiftung (Germany),
J. Bolyai Research Scholarship, EFOP, the New National Excellence
Program ({\'U}NKP), NKFIH, and OTKA (Hungary),
Department of Atomic Energy and Department of Science and Technology 
(India),
Israel Science Foundation (Israel), 
Basic Science Research Program through NRF of the Ministry of 
Education (Korea),
Physics Department, Lahore University of Management Sciences (Pakistan),
Ministry of Education and Science, Russian Academy of Sciences,
Federal Agency of Atomic Energy (Russia),
VR and Wallenberg Foundation (Sweden), 
the U.S. Civilian Research and Development Foundation for the
Independent States of the Former Soviet Union, 
the Hungarian American Enterprise Scholarship Fund,
the US-Hungarian Fulbright Foundation,
and the US-Israel Binational Science Foundation.

\end{acknowledgments}



%
 
\end{document}